%Paper: hep-th/9210144
%From: altherr@lapcls.in2p3.fr
%Date: Wed, 28 Oct 92 14:02:36 GMT

%
%     Plain TeX   (figures not included)
%
\font\titre=cmbx10 scaled\magstep2
\font\auteurs=cmr10 scaled\magstep1
\baselineskip=30pt
%--------
\vglue3cm
\line{\vbox{\halign{\hfill#\hfill\cr
Nice INLN 92/16 \cr}}
\hfill\vbox{\halign{\hfill#\hfill\cr
October 1992\cr}}}
\noindent ENSLAPP-A-372/92
\vglue3cm

\centerline {\titre THERMAL FIELD THEORY AND INFINITE STATISTICS}
\medskip
\medskip
\centerline { \auteurs Tanguy Altherr (*) and Thierry Grandou
($\dag$) }
\bigskip\medskip
\centerline {(*) Laboratoire de Physique Th\'eorique ENSLAPP}
\centerline {Chemin de Bellevue, BP110, F-74941 Annecy-le-Vieux
cedex}
\bigskip
 \centerline { ($\dag$) Institut Non Lin\'eaire de Nice UMR
CNRS 129}

\centerline {Universit\'e de Nice-Sophia Antipolis- Parc Valrose
F-06034 Nice cedex}
\bigskip
\centerline{\bf ABSTRACT} \medskip
We construct a quantum thermal field theory for scalar particles in the case of
infinite statistics. The extension is provided by working out the
Fock space realization of a ``quantum algebra", and by identifying the
hamiltonian as the energy operator.
We examine the perturbative behavior of this theory and in particular
the possible extension of the KLN theorem, and argue that it appears as a
stable structure in a quantum field theory context.
\bigskip
\centerline{(revised version)}
\vfill  \eject

\centerline {\bf 1. Introduction}
\bigskip\bigskip
Quantum Field Theories (QFT's) at finite temperature and/or densities
have been studied extensively during the past few years, and there is no need
to recall how crucial they are for our understanding of microphysics and
cosmology. However, intimately related to these physical aspects, there are
many important theoretical questions which are not fully understood.
For example, even the very existence of perturbative regimes for those
theories is not entirely obvious. Partially based on $C^*$-algebra analysis,
some authors have developed convincing arguments in favor of an inherent non
perturbativeness of QFT's at finite temperature [1]. Feynman rules can be
devised though, both in real and imaginary time formalisms [2], and the
resulting perturbative series, however formal, are certainly worth studying.
Not surprisingly, one then learns that some convenient resummations of the
original series are sometimes necessary in order that the perturbative
calculations make sense [3].

Setting aside the question of the stability of the perturbative expansions at
finite temperature [4], any such expansion has to cope with the problem of
Infra-Red (IR) singularities. At $T=0$, the infra-red character of the QFT's
perturbative series is admittedly under control: famous results such as the
Kinoshita-Lee-Nauenberg (KLN) theorem, the Poggio-Quinn theorem or the
Bloch-Nordsieck mechanism [5] set the perturbative approach on a sound basis,
at
least in this respect. Finite $T$-extensions of those results have been looked
for in the past four years and are still the subject of current
investigation [6].

The KLN theorem in particular, has been proven to hold at first [7] and
second [8] non-trivial orders of finite temperature perturbation theory, by
several authors. In the scalar case, an iterative proof for the
cancellation of the most IR singularities at all orders, has been proposed [9].

Along with these calculational proofs, a remarkable feature appeared.
Indeed, the cancellation of singularities appears to be due to ``the
miraculous properties" of the Bose and Fermi statistics. The point was
stressed by some authors [7,10], and trials performed with different
statistical distributions were shown to spoil completely the IR-finite
character of the perturbative series [11].

On the other hand there is an alternative opinion about this issue.
The overall cancellation of the IR singularities would just be the reflection
of a deeper coherence of the theory. Unitarity is usually thought of, or some
of its involved formulations [12]. But unitarity is a general principle,
and it is not always easy to handle it in the course of practical
calculations. Of course, at $T=0$, the Cutkosky rules can be regarded  as
an explicit expression for it [13]. But such an operational expression is
precisely lacking at non zero $T$ where, in spite of the ``cutting rules"
generalized by Kobes and Semenoff [14], one cannot define asymptotic
states, and thereby escape the ambiguities of a Feynman amplitude
calculation [15]. Such a situation is likely to preclude the existence of any
general argument which would decide which kind of a mechanism is responsible
for the IR-singularities cancellation, and what is its relation to
conventional Bose and Fermi statistics. \smallskip
\bigskip
In $1953$, Green observed that statistics others than Bose and Fermi
could be envisaged within the axiomatic context of QFT's [16]. This
observation has been at the origin of a rapid development of the
so-called ``parastatistical fields" [17]. These fields are quantized
according to commutation relations which are trilinear in the creation
and annihilation operators, and are characterized by an integer $p$, the
order of the parastatistical field. This number basically corresponds to
the number of quanta in a given symmetric or anti-symmetric state.

The situation referred to as ``infinite statistics" corresponds to the
case where the number $p$ is left as a free parameter without restriction,
and turns out to be realized by the so-called ``quantum" or ``deformed"
algebras. Our calculations will be performed in this framework.
\bigskip
The paper is organized as follows. In Section 2, we first recall the
basics of ``deformed quantizations" in some detail. Then we examine their
compatibility with the subsequent constraints brought about by the context of
covariant statistical mechanics. In section 3, we derive the expression for the
contour propagator of a ``$q-$deformed" scalar field. This is readily used to
set up the finite $T$-perturbation theory of both imaginary and real time
formalisms. Section 4 exhibits a direct use of these rules. A one-loop
calculation is exhibited, and the validity of KLN theorem is examined
in the case of a scalar field obeying infinite statistics.
Our comments and conclusions are presented in Section 5.
\eject

{\centerline{{\bf 2. Deformed algebras}}}
\bigskip {\bf 2.1 Deformations}\bigskip
Finite $T$ formalisms rely on a formal analogy between
imaginary time and inverse temperature. Only the time variable is involved in
this implementation of temperature, so that we can first ignore the spatial
degrees of freedom and focus on simpler quantum statistical mechanics. We
therefore consider the one dimensional harmonic oscillator. Since a free field
is nothing but a superposition of independent frequencies, generalization to
QFT will be straightforwardly obtained in the next section.
\medskip
Let us now introduce the deformed algebras. For reasons to be discussed
shortly, we choose to follow the basics of Macfarlane's presentation
[18]. We thus assume the existence of a Hilbert space ${\cal H}$ spanned by the
vector basis $\lbrace |n> \rbrace$, ${\cal H}=\bigl\lbrace |n>\ ;\ n\in I\!N
\bigr\rbrace$, and endowed with the scalar product $<m|n>=\delta_{mn},
\forall m,n$. Acting upon ${\cal H}$, we define a set of three linear
operators $a$, $a^+$ and $N$ such that
$$
       a|0>=0,\ \ \ a^+|n>={\sqrt {[n+1]}}\ |n+1>,\ \ \
       a|n>={\sqrt {[n]}}\ |n-1>,\ \ \ N|n>=n|n>
\eqno (2.1)$$
These operators will be taken in the Heisenberg picture. In what follows,
 the bracketted quantities will stand for either
$$
         [r]={1-q^r \over {1-q}}
\eqno(2.2)$$
or
$$
         [r]={q^r-q^{-r} \over {q-q^{-1}}}
\eqno(2.3)$$
Indeed we will be mostly involved in the second case $(2.3)$.
Over the representation ${\cal H}$, the deformation parameter $q$ can be
chosen so that the operators $a$, $a^+$ and $N$ satisfy
$$
             a^{\dag} = a^+,\ \ \ [N,a^+]=a^+,\ \ \ [N,a]=-a
\eqno(2.4)$$
where the symbol $\dag$ refers to the adjoint. Also, by formally
extending to operators the above brackets postulated for scalars, one gets
$$
        a^+a=[N]\ ,\ \ \ aa^+=[N+1]
\eqno(2.5)$$
Now for the postulated hermiticity of $N$ to be consistent with
properties $(2.4)$ and $(2.5)$ on the one hand, and the definition of the
brackets (2.2) and $(2.3)$ on the other hand, one must choose the
parameter $q$ a real number or a pure phase only. The same consistency argument
imposes that $q$ be real valued in the definition $(2.2)$. Then one immediately
verifies that the operators $a$, $a^+$ and $N$ satisfy
$$
         aa^+-qa^+a=1
\eqno(2.6)$$
in the case of definition $(2.2)$, and
$$
         aa^+-qa^+a=q^{-N}\ \ \
          {\rm and/or}\ \ \ aa^+-q^{-1}a^+a=q^N
\eqno(2.7) $$
in the case of definition $(2.3)$. That is, one gets two deformations,
$q$-parametrized, of the familiar commutation algebras. Eventually,
the and/or of $(2.7)$ displays the symmetry of this latter deformation
under the exchange $q\leftrightarrow q^{-1}$.
\medskip
Some remarks are in order.
 \smallskip {\bf (i)} At first sight, this construction may appear to
be a somewhat formal and indirect way to introduce the deformations.
However, it has the advantage of embedding the deformed algebras in their Fock
space realization from the onset; this is certainly of interest when one
tries to implement them as QFT's. Moreover, being a construction, it
automatically enjoys self-consistency, provided it is not an empty one. It is
not. An explicit coordinate-space realization of the $q$-deformed harmonic
oscillator has been proposed [18] in the case of definition $(2.3)$.
The operators $a$, $a^+$ and $N$ are referred to as the ``$q$-deformed"
annihilation, creation, and number operators respectively. Note also that the
hermitic conjugation of operators $a$ and $a^+$, eq.(2.4), ensures
the unitarity of the representation.
  \smallskip
 {\bf (ii)} The limit $q=1$ reproduces the original bosonic quantification for
both deformations. This is not so at the limit $q=-1$, which is not fermionic
in
the second case $(2.7)$. For this reason, an alternate deformation is
sometimes considered [19], which continuously extrapolates between both
statistics
$$  aa^+-qa^+a=q^{-2N}    \eqno(2.8)$$
We will not consider it though, for we found that its Fock space
realization encounters serious contradictions. We will therefore exclude the
case $q=-1$ and consider deformations about the bosonic statistics only [20].
\smallskip {\bf (iii)}
 The Hilbert space ${\cal H}$ is positive definite by construction
$$\forall n\in I\!N\ \ , \ |n>={(a^+)^n\over {\sqrt {[n]!}}}\ |0>\ ,\
\ \
[n]!=[n]\ [n-1]\ ..[1]\ ,\ \ \ ||\ |n>\ ||^2=1  \eqno(2.9)$$
When $q$ is a pure phase, $i.e.\ q =e^{i\theta}$,
one can parametrize $\theta=\pi r$ whith $r$ some number. If $r$ is a rational
number ($r=k/p$), then the Hilbert space is spanned by $p$ states
($|j>,j=0,1,..,p-1 $) [21]. If $r$ is not a rational number, there is an
infinite number of such states, as in the ordinary case, but there exists an
integer $n$ such that $[n]<0$ implying that $a$ and $a^+$ are no longer
hermitian conjugates (this is easily understood by inspection of eq.(2.3)).
Fortunately, most formulae turn out to be independent
of this $\theta$-parametrization and so are the results of the next sections.
\smallskip
{\bf (iv)} As compared to the ordinary Bose case, a very distinctive
feature of the deformations is worth emphasizing. Considering a system of
several degrees of freedom denoted by indices $i,j$ (continuous or discrete),
the quantizations (2.6) and (2.7) are not a priori completed by such relations
as $$
         [a_i,a_j]=[a_i^+,a_j^+]=0
\eqno(2.10)$$
In effect, their would be generalization
$$
         a_ia_j-qa_ja_i=a_i^+a_j^+-qa_j^+a_i^+=0
\eqno(2.11)$$
is easily proven to hold only if $q^2=1$, that is in the standard
cases. But, as in these standard cases, such relations are not necessary
to define operator's normal ordering, Wick's theorem and the calculation of
matrix elements. If we choose not to impose the set of relations (2.10),
the unusual consequence is that the $n!$ permutations of an
$n$-identical particles state, now become $n!$ linearly independent states
[22].

\bigskip The ``$q$-deformed harmonic oscillator" is accordingly defined in
terms
of the above deformed operators. Within familiar conventions for the
constants, one has $$
      H={1\over 2m}\ P^2+{1\over 2}m\omega^2\ X^2
\eqno(2.12)$$
with
$$
        X={\sqrt{\hbar \over {2m \omega}}} (a+a^+)\ \ \ {\rm and}\ \
\
        P=i{\sqrt{\hbar m \omega \over 2}}(a-a^+)
\eqno(2.13)$$
and finds
$$
        H={1\over 2}\hbar \omega\left([N]+[N+1]\right)
\eqno(2.14)$$
In terms of $a$ and $a^+$, the number operator $N$ can be given the
formal expressions $$
  N={\ln\ \left([a,a^+]\right)\over \ln\  q}
      \ \ \ \ \ \ {\rm and}\ \ \ \ \ \
  N=-{\ln\ \left([a,a^+]_q\right)\over \ln\  q}
\eqno(2.15)$$
for the deformations $(2.6)$ and $(2.7)$ respectively,
whereas in the second case we have introduced the ``$q-$mutator"
$$
       [a,a^+]_q\equiv aa^+-qa^+a
\eqno(2.16)$$
Both formal expressions reduce to the usual number operator $N=a^+a$ in
the limit $q=1$, and can be defined as infinite series expansions in terms of
deformed operators $a^+$ and $a$ [19],[22].
\bigskip
\medskip {\bf 2.2 Thermalization}
\bigskip
 At thermal equilibrium, statistical averages
typically involve quantities of the kind
$$
{\cal N}\  {\rm Tr}\left(e^{-\beta H}\ {\rm T}\ Pol \lbrace
X(t)\rbrace \right)
\eqno(2.17)$$
where ${\cal N}$ is some relevant normalization, $\beta$ the
inverse temperature, $Pol \lbrace X(t)\rbrace$ some polynomial in the position
operators and T a prescription of time ordering. The point is that a trace
must be taken over a complete set of states. For the calculations to be
meaningful, it is therefore crucial that the cyclicity of the trace
be consistent with the underlying algebraic structure.

\medskip
For short, we hereafter set $\hbar =1$ everywhere.
For operators in the Heisenberg picture, the equation of motion reads
$$
              {{\rm d}A\over {\rm d}t}= -i[A,H]
\eqno(2.18)$$
and it is important to realize that the cyclicity of the trace
precludes any $q$-extension of the above equation, where for instance
the usual right hand side commutator would be replaced by the corresponding
$q$-mutator $[A,H]_q$.\medskip

The Hamiltonian usually dealt with is given by (2.14) [23,24]. As our
Hamiltonian though, we will rather take
$$
            H=\omega (N+C^{te})
\eqno(2.19)$$
with $N$ the deformed number operator of $(2.15)$, and
thereby identify the Hamiltonian with the energy operator. One
recovers the well known relations
$$
            [H,a]=-\omega a\ \ \ \ {\rm and}\ \ \ \
            [H,a^+]=\omega a^+
\eqno(2.20)$$
and two properties follow by construction: the positive
definiteness of the energy eigenvalues for all $n$ and $q$-values, and, most
importantly, the possibility of well defined statistical averages.
Considering for example the two-point function
$G(t)\equiv {\cal N}\  {\rm Tr}\left(e^{-\beta H}\ {\rm T}\ X(t)X(0) \right)$,
we have
$$
        G^+(t-i\beta) = G^-(t)
\eqno(2.21)$$
where $G^{\pm}$ refers to the usual decomposition
$$
        G(t)=\Theta(t)G^+(t)+\Theta(-t)G^-(t)
\eqno(2.22)$$
that is, the Kubo-Martin-Schwinger (KMS) conditions [2] are satisfied
by the two-point function at thermal equilibrium. One can check
easily that these two crucial properties are not clearly at issue if the
Hamiltonian (2.14) is used instead. Note that the limiting value of
the constant appearing in (2.19) is known only when $q\rightarrow 1$.
However this latter non-determination will never show up in the subsequent
analysis, so that we can either take $C^{te}=1/2$ or simply drop it. Note also
that this choice of $H$ can be compared with the ones performed in
parastatistics [25] and is in full agreement with the traditional choice of an
Hamiltonian for a fermionic oscillator [26].

\bigskip\medskip
{\bf 2.3 Deformed statistics}
\bigskip
It is now an easy task to calculate the deformed Bose-Einstein
distributions. Considering the deformation $(2.7)$, one calculates the
thermal averages of both sides, that is
$$
        <aa^+>_\beta-q<a^+a>_\beta =<q^{-N}>_\beta
\eqno(2.23)$$
Within obvious notations one has
$$
        \eqalignno{ <q^{-N}>_\beta &={1\over Z}\sum <n|e^{-\beta H}\
        e^{-iN\theta}|n>\cr &= {1\over Z}\ e^{-{\beta\omega}/2}\
        {1\over 1-q^*e^{-\beta \omega}}
&(2.24)\cr}$$
By evaluating the partition function $Z$ we eventually get
$$
        <q^{-N}>_\beta ={{e^{\beta \omega}-1}\over {e^{\beta\omega}-q^*}}
\eqno(2.25)$$
Then by using the commutators $(2.20)$ and the cyclicity of the
trace, one derives
$$
          <q^{-N}>_\beta =\left({e^{\beta \omega}-q}\right)\ <a^+a>_\beta
\eqno(2.26)$$
and eventually
$$
           <a^+a>_\beta\  =\ { {e^{\beta \omega}-1}\over
         {{e^{2\beta \omega}-2\cos\theta e^{\beta\omega}+1}}}
\eqno(2.27)$$
that is, a real quantity for all $\theta$.
Following the same steps, the first algebra leads to a simpler expression
$$
           <a^+a>_\beta\  =\ { 1 \over e^{\beta\omega} - q}
\eqno(2.28)$$
Then we can analyze the generalization to free quantum field theory.

\bigskip\bigskip
{\centerline {\bf 3. Perturbation Theory}}
\bigskip
{\bf 3.1 Locality}
\bigskip
We now consider the extension to free scalar field
theory, and begin by briefly reviewing the possibility of
a local realization of the deformations. Locality is
presumably one of the most tricky points when one wishes to implement the
deformations as (local) QFT's. This axiom is referred to as the basic
requirement that observables be pointlike functionals of the fields, and
commute at spacelike separations [27]. Indeed, a general theorem [27]
prevents any field theory with statistics other than Fermi/Bose or
parafermi/parabose, from being realized as local QFT's (the latter
``para"-alternative being restricted to $D\geq 3$ dimensional spaces).
Even though, TCP theorem has been shown to hold true for  ``$q$- deformed"
free fields [22].
\smallskip
When $q-$deformed fields are considered, locality
considerations are usually made on the basis of $q-$mutators, which then play
the role of ordinary commutators. This generalization can be thought of as
being
natural, and is intended to take advantage of the $q-$mutators
 $$
     a(k)a^+(p)-qa^+(p)a(k)=(2\pi)^3 2\omega_k \ q^{-N_k} \ \delta^{(3)}(k-p)
\eqno(3.1)$$
But, considering such quantities as $[\phi(x),\phi(y)]_q$, implicitly
supposes that the underlying space-time manifold is endowed with some
non commutative geometrical properties [28]. Whether it exists or not any
compelling reason to consider such situations, falls beyond the scope of
the present analysis. Indeed, one can check that local $q-$mutativity is not
satisfied  with the deformations (2.6) and (2.7), and we will
restrict ourselves to ordinary locality considerations.
\medskip
Take the free hermitian scalar field $\phi(x)$ in a four dimensional
space-time
$$
     \phi(x)=\int {{\rm d}^3k \over {(2\pi)^3\ 2\omega_k }}
             \left( a(k) e^{-ik.x}+a^+(k) e^{ik.x}\right)
\eqno(3.2)$$
where the Fourier components are quantized according to the deformation
$(3.1)$, and act linearly on the Fock space of states obtained by the
standard procedure (smeared polynomials of $a^+(k)$ acting upon the vacuum
[22]).
\medskip
Considering the commutator of two free field operators at space-like
separation, one finds, as expected in view of the theorem cited above
$$
        [\phi(x),\phi(y)]_{x_0=y_0} \neq 0
\eqno(3.3)$$
However, the thermal average of the commutator reads
$$
       <[\phi(x),\phi(y)]>_\beta =\int {{\rm d}^4k \over {(2\pi)^4}}
       e^{-ik.(x-y)}2\pi\epsilon(k_0)\delta(k^2-m^2)
       n_\theta(\omega_k) (e^{\beta |k_0|} - 1)
\eqno(3.4)$$
where $\epsilon (k_0)$ is the usual sign distribution, and where we have
introduced the notation
$$
   n_\theta(\omega_k)\equiv <a^+_ka_k>_\beta ={ {e^{\beta\omega_k}-1}
   \over{{e^{2\beta\omega_k}-2\cos\theta e^{\beta\omega_k}+1}}}
\eqno(3.5)$$
At space-like separations one gets
$$
   \left(\ <[\phi(x),\phi(y)]>_\beta \right) |_{x_0=y_0} =0
\eqno(3.6)$$
that is, local commutativity is restored by taking the statistical
average, which is of course pretty remarquable and unexpected a result.
This important issue is certainly of interest in order to get a
reliable perturbation theory, as we will observe in the following.

\vfill\eject
{\bf 3.2 The contour propagator}

 \bigskip Let $G_C(x-y)$ be the contour propagator
$$
     G_C(x-y) =<T_C\ \phi(x)\phi(y)>_\beta
\eqno(3.7)$$
where the mean value refers to the thermal average $(2.17)$, whereas the
prescription $T_C$ orders the field operators along some given contour
$C$ of the time-complex-plane. Defining the $G^\pm$ components by
$$
       G_C(x-y)=\Theta_C(x_0-y_0)\ G^+(x-y) + \Theta_C(y_0-x_0)\ G^-(x-y)
\eqno(3.8)$$
one has (see (2.21))
$$
      G^+(x-y-i\beta)= G^-(x-y)
\eqno(3.9)$$
that is the KMS condition. Taking equation $(3.2)$ into
account, $G_C$ can be explicitly written as
$$
   \displaylines {G_C(x-y)=\int {{\rm d}^3k \over {(2\pi)^3
   \ 2\omega_k}}\Bigl\lbrace \Theta_C(x_0-y_0)\left(<aa^+>_\beta
   e^{-ik(x-y)}+<a^+a>_\beta e^{ik(x-y)} \right) \hfill \cr \hfill +
   \Theta_C(y_0-x_0)\left(<a^+a>_\beta e^{-ik(x-y)}+<aa^+>_\beta
   e^{ik(x-y)}\right)\Bigr\rbrace \qquad\qquad
(3.10)}$$
Once a contour is given, it is just a book-keeping device to deduce
the corresponding set of Feynman rules.
\bigskip
\medskip{\bf 1) The real time formalism}
\medskip We will consider the oriented contour which is usually
choosen, depicted in Fig.1, and basically composed of the real axis and of
its $-i\sigma$-shifted part, counter-oriented [2]. The parameter $\sigma$ is
taken in the range $0\leq \sigma \leq \beta$. The real time free propagator is
a two by two matrix with components $G^{rs},\ \lbrace r,s \rbrace =1,2$.

\smallskip
By using $(3.10)$, one gets for temporal arguments lying on the
$C_1$ part of $C$
$$
      G^{11}(x-y)=\int {{\rm d}^4k\over (2\pi)^4}\ e^{-ik(x-y)}
            \bigl\lbrace <aa^+>_\beta\ \Delta(k)+<a^+a>_\beta\
            \Delta^*(k)\bigr\rbrace
\eqno(3.11)$$
where $\Delta(k)$ is the usual Feynman propagator
$$
           \Delta(k)={i\over {k^2-m^2+i\epsilon}}
\eqno(3.12)$$
Then, for the $C_2$ part, one has
$$
       G^{22}(k)=\left(G^{11}(k)\right)^*=\ <aa^+>_\beta\
                 \Delta^*(k)\ +<a^+a>_\beta\ \Delta(k)
\eqno(3.13)$$
Non diagonal components are those for which $x_0$ belongs to
$C_1$, $y_0-i\sigma$ to $C_2$, and vice versa
$$
\eqalignno{G^{12}(x-y) &=\int {{\rm d}^3k \over {(2\pi)^3 2\omega_k}}
\Bigl\lbrace
n_\theta(\omega)e^{\omega\sigma}e^{-ik.(x-y)}+e^{\beta\omega}n_\theta
(\omega)
e^{-\omega\sigma}e^{ik.(x-y)}\Bigr\rbrace \cr &=\int {{\rm d}^4k\over
{(2\pi)^4}}e^{-ik.(x-y)}\Bigl\lbrace 2\pi\delta(k^2-m^2)e^{\sigma
k_0}n_\theta(\omega)\left(\Theta(k_0)e^{\beta\omega} +
\Theta(-k_0) \right)\Bigr\rbrace &(3.14)\cr}$$
where we have used the relations $(2.23)$ and $(2.26)$ in
order to relate $<aa^+>_\beta $ to $<a^+a>_\beta $. A similar expression can be
derived for the $G^{21}$ component. In momentum space, the result can be
summarized as
$$
        G^{12}(k)=2\pi\delta (k^2-m^2)n_\theta(\omega)e^{\sigma
        k_0}\left(\Theta(k_0)+\Theta(-k_0)e^{\beta\omega}\right)
\eqno(3.15)$$
and likewise
$$
         G^{21}(k)=2\pi\delta (k^2-m^2)n_\theta(\omega)e^{-\sigma
        k_0}\left(\Theta(-k_0)+\Theta(k_0)e^{\beta\omega}\right)
\eqno(3.16)$$
Thanks to energy conservation, the calculation of Green's functions
is independent of $\sigma$ [2], and a convenient choice is the one performed
in Thermo Field Dynamics, which sets $\sigma =\beta /2$. One finds
$$
       G^{12}(k)=G^{21}(k)=2\pi\delta(k^2-m^2)
                 e^{\beta\omega/2}n_\theta(\omega)
\eqno(3.17)$$
completed by
$$
       G^{11}(k)=\left(G^{22}(k)\right)^*=n_\theta(\omega)\left(\Delta(k)
                 e^{\beta\omega}+\Delta^*(k)\right)
\eqno(3.18)$$
The two previous equations set up the Feynman rules for a free scalar
$q-$deformed field, assuming that the bare vertices remain unaltered by
the deformation, that is
$$
          (-ig)\ {\rm for\ type\  1\ vertices\ ,
  \ and}\ (+ig)\ {\rm for\ type\ 2\ vertices}
\eqno(3.19)$$
and no couplings between the different types of fields ($g$ stands
for the bare coupling constant).

\medskip
By the same token, the cut propagators $G^{\pm}$ which are conveniently used
for the calculation of imaginary parts of Green's functions [13,14], are easily
derived by setting $\sigma=0$ in $(3.15)$ and $(3.16)$, [10]. One finds
$$
       G^-(k)=G^{12}(k)\bigr\rbrack_{\sigma=0}=\left(\Theta(k_0)+
       \Theta(-k_0)e^{\beta\omega}\right)n_\theta(\omega)\
       2\pi\delta(k^2-m^2)
\eqno(3.20)$$
$$
       G^+(k)=G^{21}(k)\bigr\rbrack_{\sigma=0}=\left(\Theta(-k_0)+
       \Theta(k_0)e^{\beta\omega}\right)n_\theta(\omega)\
       2\pi\delta(k^2-m^2)
\eqno(3.21)$$

\medskip
Each of these involved expressions can be shown to reduce to the well-known
ones in the limit of no deformation, $\theta=0$. For later purposes
we here give these limits for the two previous equations
$$\eqalignno{lim_{\
\theta =0}\ G^{\pm} (k) &=\left(\Theta(\mp k_0)+ \Theta(\pm
k_0)e^{\beta\omega}\right)n_B(\omega)\ 2\pi \delta(k^2-m^2)\cr
&=2\pi\left(\Theta(\pm k_0) +n_B(\omega)\right)\delta(k^2-m^2)
&(3.22)\cr}$$
where the first expression of the RHS of (3.22) will consequently
be considered as more fundamental than the second customary one, which is
specific to the pure bosonic case only. This observation will play a central
role in the next section. Another way to understand this, is to note that a
direct replacement of $n_B$ by $n_\theta$ in the usual real-time Feynman rules
would yield an incorrect result. We will similarly make use of the limit of
(3.18)
$$\eqalignno{lim_{\ \theta =0}\ G^{11} (k) &=n_B(\omega)\left(
  e^{\beta\omega}\ \Delta(k)+\Delta^*(k)\right) \cr &=\left((1
 +n_B(\omega))\Delta(k)+n_B(\omega)\Delta^*(k)\right)
&(3.23)\cr}$$

 \medskip Now let us come to the following striking feature: considering the
zero temperature limit of the above Feynman rules, one gets the usual
$T=0$ limit of an undeformed theory, that is
$$
    lim_{\ T=0}\ G^{11}(k)=\Delta(k)\ ,\ \ \ lim_{\ T=0}\
     G^{22}(k)=\Delta^*(k)
\eqno(3.24)$$
and likewise, for the cut propagators
$$
            lim_{\ T=0}\ G^{\pm}(k)=2\pi\Theta(\pm k_0)\delta(k^2-m^2)
\eqno(3.25)$$
Eventually,
$$
         lim_{\ T=0}\ G^{12}(k)=lim_{\ T=0}\ G^{21}(k)=0\ ,\forall \theta
\eqno(3.26)$$
All the effects of the deformation have disappeared in the
limit of a zero temperature. This can be readily realized by taking the
infinite $\beta$ limit in an expression such as $(3.10)$, where only the
vacuum state is seen to survive. This rather peculiar feature has recently
been noticed [29]. One may note, however, that the deformation is manifest
at the 4-point function level [22].

\medskip
The set of Feynman rules explicited above define a well behaved
perturbation theory for free fields that we shall use in the following.
Because of equations
(3.24)-(3.26), the ultra-violet sector of this theory is unaltered by the
deformation. This makes it possible to renormalize the theory at $T=0$, as is
usually done in the undeformed case.  Note that the infra-red sector is
modified
instead. In effect, the customary IR behavior of the Bose distribution
$$
     n_B(\omega) \sim {1\over \beta\omega}\ ,\ \ \ \beta\omega\ll 1
\eqno(3.27)$$
is now replaced by
$$
       n_\theta(\omega) \sim {\beta\omega \over{2(1-\cos \theta)}} \
        ,\ \ \ \beta\omega\ll 1
\eqno(3.28)$$
Thus the change is drastic : note the gap of two powers of
$(\beta\omega)$ between $(3.27)$ and $(3.28)$, as soon as the
deformation is switched on. Had we used the deformation $(2.6)$ instead, a gap
of one power of $\beta\omega$ had been obtained.

For the sake of completeness, we next give the free deformed propagator in the
imaginary time formalism.

\bigskip\bigskip\bigskip {\bf 2) The imaginary time formalism}
\bigskip
 This formalism is generated by the simplest
choice of a contour, namely a straight line stretching from $t=0$ to
$t=-i\beta$ on the imaginary time axis. The $T_C$ ordering reduces to the
so-called temperature ordering $T_\tau$, where the variable $\tau=it$ can be
restricted to the interval $-\beta \leq \tau \leq \beta$, in view of
periodicity. Taking relations $(2.23)$ and $(2.26)$ into account, one
straightforwardly derives the imaginary time deformed propagator in the
popular form
$$
       \Delta_\theta (\tau,k)={1\over 2\omega}\left(\  \
      e^{\omega (\beta - |\tau|)}+\ e^{\omega |\tau|}\
     \right)n_\theta (\omega)
\eqno(3.29)$$
This propagator can be checked to verify the periodicity conditions
induced by the KMS condition, that is, one has $\Delta_\theta (\tau -\beta,k)
=\Delta_\theta (\tau,k)$  for $0\leq \tau \leq \beta$, and
$\Delta_\theta (\tau +\beta,k)=\Delta_\theta(\tau,k) $, for
$-\beta \leq \tau \leq 0$.

 \bigskip\bigskip\bigskip {\centerline {\bf 4. The KLN theorem}}
\bigskip
We now come to our original goal. As advertised in
the introduction, the thermal extension of the KLN theorem seems due to
some ``miraculous properties" of the Bose and Fermi distributions. In
effect, for a large variety of three body processes (where a particle with
energy $y$ splits into two others with energy $x$ and $y-x$), relations such as
[7]
$$
     n_B(x)\ n_B(y-x)\ =\ n_B(y)\ (\ 1+n_B(x) + n_B(y-x)\ )
\eqno(4.1)$$
or in QED and QCD, such as [7,10]
$$
     n_B(x)\ n_F(y-x)=n_F(y)\ (\ 1+n_B(x)-n_F(y-x)\ )
\eqno(4.2)$$
have been recognized to be at the
origin of the cancellation of IR (including collinear) singularities. Thanks to
these relations, real singular contributions add up to match
singular virtual ones. Restricting ourselves to the bosonic case, it can be
shown that all these ``miraculous properties" derive from the only basic
relation
$$
      n_B(x+y)\ \left(1+n_B(x)\right)\ \left(1+n_B(y)\right)\
      =\ \left(1+n_B(x+y)\right)\ n_B(x)\ n_B(y)
\eqno(4.3)$$
which will be interpreted shortly. Now, relation (4.3) is not
satisfied by the deformed statistical distribution $n_\theta$, so that one
could think that the cancellation of IR singularities is jeopardized. But
indeed, (4.3) turns out to be specific to the case of zero deformation,
$i.e.$ the pure bosonic case, and this explains why trials performed with
different distributions were doomed to failure.

However, in the case of non zero deformation one has
$$
         <a^+a>_\beta (x+y)\ <aa^+>_\beta (x)\ <aa^+>_\beta (y)\
      =\ <aa^+>_\beta (x+y)\ <a^+a>_\beta (x)\ <a^+a>_\beta (y)
\eqno(4.4)$$
that is, in view of (2.23) and (2.26)
$$
    n_\theta (x+y)\ \ e^{\beta x}n_\theta (x)\ \ e^{\beta y}n_\theta (y)\
  =\ e^{\beta (x+y)}n_\theta (x+y)\ n_\theta (x)\ n_\theta (y)
\eqno(4.5)$$
By observing that  $e^{\beta x} n_B(x)=1+n_B(x)$  one realizes that the
``miracle'' (4.3) falls into a whole, much wider, family of similar
$\theta$-indiced ``miracles" (4.5). Indeed, the formal triviality of these
relations reveals their conservation law character. In effect, by consistently
associating the weight $n_\theta(x)$ to an incoming particle, and the weight
$e^{\beta x}n_\theta(x)$ to an outgoing one, relations (4.3) and (4.5) are
readily recognized as expressing the well-known micro-reversibility property
of elementary processes. As such, they extend to many more kinds of elementary
processes (they are pictured in Fig.2 for three body ones). Also, they display
the crucial role played by the KMS condition in this respect. In its turn, the
KMS condition is rendered possible by choosing (2.19) as the Hamiltonian.
\medskip\smallskip
Property (4.5) is not sufficient to ensure that KLN theorem can be
satisfied in the deformed case. In this latter situation, one also has to make
sure that perturbative calculations develop analytic structures which are the
same as the ones at play in the ordinary case of no deformation. We therefore
turn to this task now.
\bigskip
Let us consider the kind of processes which have
been studied till recently [7,8] , that is the damping rate of a heavy photon
or a Higgs particle in a thermal bath. We assume that the Higgs can decay
into scalar particles which have self-cubic interactions. Since these are
dynamical quantities, the real time formalism is more appropriate. At first non
trivial order of perturbation theory, this rate is given by the imaginary part
of the self-energy function of the particle. We restrict ourselves to the
topology of Fig.3 , as the cancellation of singularities is known to take place
in each topology separately [5].\medskip The best way to proceed consists in
briefly recalling the structures involved in the standard situation. The
thermal
free propagator can be conveniently written as
$$
       G(k) = {\cal U}\pmatrix{{\Delta(k)} & {0}\cr
                                       {0} & {\Delta^*(k)}\cr}{\cal U}
\eqno (4.6)$$
where ${\cal U}$ stands for the unimodular matrix
representation of a  Bogoliubov transformation (not unitary!) [2]
$$
        {\cal U} =\pmatrix {\sqrt{1 +
         n_B(k_0)}&\sqrt{n_B(k_0)}\cr\sqrt{n_B(k_0)}&\sqrt{1 + n_B(k_0)}\cr}
\eqno(4.7)$$
Assuming a Schwinger-Dyson equation
for the dressed propagator ${\cal G}^{rs}(k)$
$$
     {\cal G }^{rs}(k)=  G^{rs}(k)+ G^{rr'}(k)\left(-i\Sigma^{r's'}(k)\right)
                         {\cal G}^{s's}(k)
\eqno (4.8)$$
one easily derives the set of relations
$$
     \left( {\cal G }^{rs}(k)\right) = {\cal U} \pmatrix{{\widehat
     \Delta(k)}&{0}\cr{0}& {\widehat \Delta^*(k)}\cr} {\cal U}
\eqno(4.9)$$
with
$$
   \widehat \Delta(k)={i\over {k^2-m^2-\widehat\Sigma(k)}}
\eqno(4.10)$$
and where the self energy function $\widehat\Sigma(k)$ is related to
the matrix self energy by the relation
$$
    \left(-i\Sigma(k)\right)={\cal U}^{-1}
    \pmatrix{{-i\widehat \Sigma(k) }&{0}\cr{0}& {+i\widehat
    \Sigma^*(k)}\cr} {\cal U}^{-1}
\eqno(4.11)$$
that is, for practical purposes
$$
    \widehat\Sigma(k)= Re\Sigma^{11}(k)
    +i\tanh(\beta k_0/2) Im\Sigma^{11}(k)
\eqno (4.12)$$
Actually the quantities $\widehat\Sigma(k)$ and $\widehat
\Delta(k)$ are those related to the corresponding
imaginary time ones, by analytic continuation. The cut propagator
with one loop self-energy correction immediately follows [8]
$$
     {\cal G}^+(k)=2e^{\beta k_0/2}\ n_B(k_0){\sqrt
     {\left(\Theta(-k_0)+e^{\beta\omega}\Theta(k_0) \right)}}\ Re \left(
     -i\widehat\Sigma(k)\ \Delta^2(k)\right)
\eqno(4.13)$$
In terms of the previously defined expressions, the relevant quantity
is conveniently expressed as
$$
     \eqalignno{\Gamma(q) &=\int {{\rm d}^Dk\over {(2\pi)^D}}\
                       {\cal G}^+(k)\ G^-(k-q) \cr
                     &=-2\int {{\rm d}^Dk\over {(2\pi)^D}}G^-(k-q)
\left(\Theta(-k_0)+e^{\beta\omega}\Theta(k_0)\right)n_B(k_0))
Re \left( -i\widehat\Sigma(k)\ \Delta^2(k)\right)
&(4.14)\cr}$$
In (4.14), $q=(q,\vec 0)$, stands for the $D$-moment of the Higgs
particle considered. The mass $m$ (and the coupling constant $g$
in $\widehat\Sigma $) can be thought of as the parameters of the $T=0$
renormalization procedure, and as we have pointed out
in Section 3.2, the same steps can be taken in the deformed case.

Now, the structure displayed by equation (4.14) has been proven to yield
IR-singularity cancellation, either by direct computation [8], using
different types of regularization, or by appealing to dispersion
relations [9].   \bigskip

In the case of deformed Feynman rules (3.17)-(3.21), there is no
unimodular Bogoliubov matrix such as (4.7). Nevertheless one
can still write
$$ G_\theta(k) = {\cal U_\theta}
         \pmatrix{ \Delta(k) & 0   \cr
                   0 & \Delta^*(k) \cr} {\cal U_\theta}
\eqno (4.15)$$
with ${\cal U_\theta}$ the invertible matrix
$$
{\cal U_\theta} =
\pmatrix{ \sqrt{e^{\beta\omega}n_\theta(k)} & \sqrt{n_\theta(k)} \cr
          \sqrt{n_\theta(k)} & \sqrt{e^{\beta\omega} n_\theta(k)}\cr}
\ \ \ \  ,\ \ \ \ \ {\rm det}\ {\cal U_\theta}
                ={n_\theta(\omega)\over n_B(\omega)}
                = {n_\theta(\omega)\over n_{\theta=0}(\omega)}
\eqno(4.16)$$
and where we hereafter append a subscript $\theta$ for the deformed quantities.

As they stand, the Feynman rules of the previous section can be used to
calculate the one-loop dressed propagator ${\cal G}_\theta^{rs}(k)$, which
is obtained by summing over the thermal indices, the one-loop self
energy insertion
$$
{\cal G}_\theta^{rs}(k)=G^{rs}_\theta (k)+G^{rr'}_\theta (k)\left
(-i\Sigma_\theta^{r's'}(k)\right )G^{s's}_\theta (k)
\eqno(4.17)$$
For the self energy matrix itself, the following relations are easily
established at one-loop order ($O(g^2)$)
$$\eqalign{
     \Sigma^{22}_\theta(k) &= -(\Sigma^{11}_\theta(k))^*  \cr
     \Sigma^{12}_\theta(k) &= \Sigma^{21}_\theta(k)  \cr
  Im \Sigma^{11}_\theta(k) &= i\cosh(\beta k_0/2) \Sigma^{12}_\theta (k)}
\eqno(4.18)$$
and at least at this order, just correspond to the standard non
deformed expressions with the distribution $n_\theta$ replacing the standard
one $n_B$. Defining the self energy function $\widehat \Sigma_\theta$ by
the relation
$$
    \widehat\Sigma_\theta(k)= {n_\theta(\omega)\over n_B(\omega)}
    \Bigl\lbrace Re\Sigma_\theta^{11}(k) +i\tanh(\beta k_0/2)
                 Im\Sigma_\theta^{11}(k)\Bigr\rbrace
\eqno(4.19)$$
the above set of relations leads to the matricial form
$$
     \left(-i\Sigma_\theta (k)\right )={\cal U_\theta}^{-1}
     \pmatrix{{-i\widehat
     \Sigma_\theta(k) }&{0}\cr{0}& {+i\widehat \Sigma_\theta^*(k)}\cr}
     {\cal U_\theta}^{-1}
\eqno(4.20)$$
with ${\cal U_\theta}$ as defined in (4.16). Note that the previous remark
does not extend to the function $\widehat \Sigma_\theta$ which is not obtained
out of (4.12), by the replacement of $n_B$ by $n_\theta$. In effect, the non
unimodular character of ${\cal U_\theta}$ induces some overall rescaling
function $(n_\theta /n_B)$. However this is a perfectly regular function of
$k$,
which factors out the singular structures involved in the one-loop series.
Accordingly, the cut propagator with one loop self energy insertion can be
written as
 $$\eqalignno{
     {\cal G}_\theta^+(k) &=2e^{\beta (k_0+\omega)/2}\
n_\theta(\omega)\
     Re \left( -i\widehat\Sigma_\theta(k)\ \Delta^2(k)\right)\cr
     &=2\left(\Theta(-k_0)+e^{\beta\omega}\Theta(k_0)\right)\
     n_\theta(\omega)\ Re \left( -i\widehat\Sigma_\theta(k)\
\Delta^2(k)\right)
&(4.21)\cr}$$
and the decay rate under consideration as
$$\eqalignno{
     \Gamma_\theta(q) &=\int {{\rm d}^Dk\over {(2\pi)^D}}\
     {\cal G_\theta}^+(k)\ G_\theta^-(k-q) \cr &=-2\int {{\rm
d}^Dk\over
     {(2\pi)^D}}G_\theta^-(k-q)\left(\Theta(-k_0)+e^{\beta\omega}\Theta(k_0)
     \right)n_\theta(k_0))
      Re \left( -i\widehat\Sigma_\theta(k)\ \Delta^2(k)\right)
&(4.22)\cr}$$
It is here made clear that, for the replacement of (standard) $n_B$
by (the deformed) $n_\theta$, the structure of the deformed case as
displayed in (4.22), is exactly the same as the one displayed in (4.14) in the
ordinary case of no deformation. Then, because of this complete structural
analogy and thanks to identity (4.5), we are now in a position to conclude
that cancellation of one-loop IR singularities will take place with deformed
statistics if it so with standard ones. In other words, one-loop KLN theorem
is satisfied with deformed statistical distributions.

For the sake of illustation, let us give some calculational details and remarks
about the cancellation of IR singularities (here collinear only). We need the
following expressions $$
        Re G^{11}(k)=\left( e^{\beta\omega}+1\right)\
                         n_\theta(\omega)\pi\delta(k^2)
\eqno(4.23)$$
$$
        Im G^{11}(k)=\left( e^{\beta\omega}-1\right)\  n_\theta(\omega)
                     \ {\cal P}\left({1\over k^2}\right)
\eqno(4.24)$$
which are easily derived from (3.18) in the massless limit. The
zero-temperature part is exactly the same as in the usual $(\phi^3)_6$ theory,
and contains all of the ultra-violet singularities. The latter are disposed of
by renormalization, and by analytic continuation from $\epsilon < 0 $ to
$\epsilon > 0 $ [8]. For the finite temperature part, we find the leading term
$$
         Re\Sigma^{11}_\theta(k)=m^2_\theta\ {k^2\omega \over K^3}\
                            \ln{\omega -K\over \omega +K}
\eqno(4.25)$$
where $K=|{\vec k}|$, and where we have introduced the finite temperature mass
$$
       m^2_\theta={g^2T^2 \over 32\pi^3}\ \int_0^\infty x{\rm d}x
                \left(\ (e^{2x}-1)n^2_\theta(x)-1\right)
\eqno(4.26)$$
Note that all of the $\theta$-dependence is being kept in the overall constant
$m_\theta$. Indeed, (4.25) is nothing but the ``Hard Thermal Loop" result
for the self-energy [3]. These HTL are endowed with remarquable properties and
it is very encouraging to see that the same structures do appear also in the
case of infinite statistics. In (4.22) one can decompose
$$
    Re\left(-i\widehat\Sigma_\theta(k)\Delta^2(k)\right)=-\pi\delta'(k^2)Re
  \widehat\Sigma_\theta(k)-{\cal P}\left({1\over (k^2)^2}\right)\ Im
\widehat\Sigma_\theta(k)
\eqno(4.27)$$
where the symbol ${\cal P}$ stands for a principal
value prescription. As usual, the potential singularities arise when expanding
around the mass-shell. In (4.25), the momentum dependence is exactly the same
as
in the non-deformed case, and therefore the analytic properties of the
self-energy at one loop order are identical to the one studied for example in
[8]. The ${\rm ln}(\omega-K)$ term in $Re\widehat\Sigma_\theta(k)$, when
dimensionally regularized, is responsible for a collinear singularity
$(1/\epsilon)$ in the wave function renormalization constant. On the other
hand,
the imaginary part of $\widehat\Sigma_\theta(k)$ which can be extracted from
(4.25) with the analytic continuation ${\rm ln}{(\omega -K)/ (\omega
+K)}\rightarrow -i\pi$, for space-like $k^2$, together with the $1/(k^2)^2$
principal part in (4.27) leads to the same diverging behavior [8]. According to
the analysis done in [9], the dispersive properties of the self-energy lead to
the cancellation of this IR singularity (here a collinear one), which therefore
verifies the KLN theorem at this order.

The same analysis can be carried through with the two-loop topology
of ref.[8]. Though lengthy, one can check that the IR most singular terms
($O(1/ \epsilon^2)$, with $D=6+2\epsilon$) cancel out, following the same
patterns as in the non deformed case. It is therefore tempting to  speculate
that our result is valid at any higher number of loops. However, this possible
extension of our result would be somewhat formal and premature. One has to make
sure that Feynman rules as defined in this paper effectively correspond to the
perturbation theory allowing for the calculation of the thermal
Green's functions of eq.(2.17). Clearly, more analysis is needed,
involving higher $n\ge 4$-point Green's functions [30].

\bigskip\bigskip\bigskip
{\centerline {\bf 5. Conclusion}}
\bigskip
By working out the QFT representation of some ``deformed" or ``quantum"
algebra, we have been able to generalize the usual bosonic Feynman rules of
finite temperature QFT, to the case of so-called ``infinite statistics", in
both real and imaginary-time formalisms. The Feynman rules derived in this
paper, are those for a scalar neutral field in arbitrary space-time
dimensions; they might be very different in other cases, the charged scalar
one included. In agreement with a general theorem, the QFT representation of
the deformation could not be made local at the operatorial level.
But the remarquable point is that locality is restored by
thermal averaging, which property let room for a well behaved
perturbation theory, at least at the two-point function level.

Indeed, the resulting perturbation theory has been shown to display the
same overall structure as the ordinary non
deformed one. The ultra-violet sector is found to be totally unaffected by the
deformation, allowing for customary zero temperature renormalization
algorithms, whereas, on the contrary, the IR sector is found to be deeply
modified. Even though, the KLN theorem which rules the
cancellation of IR perturbative singularities, has been shown to hold
in case of deformations, at first non-trivial orders of perturbation theory,
if it is satisfied in the corresponding non-deformed situations.\medskip

In particular, this one-loop verification of KLN has proven that the fine
tuning mechanism responsible for IR singularity cancellation, is not due to
some ``miraculous properties" of the original Bose-Einstein statistical
distribution, as one might have thought at first. The ``miraculous
properties" themselves have been seen to fall into a much wider family of
similar properties, which all express the micro-reversibility property
of elementary processes. In this respect, the KMS conditions have been
recognized as playing a crucial role in allowing for micro-reversibility, and
we think of this latter property as one of the basic conditions ensuring the
validity of KLN.

The KLN theorem was originally proven in Quantum Mechanics and in QFT at
zero temperature. Recently, enough evidence has been accumulated in favour
of its validity at non-zero temperature and/or chemical potential too.
Throughout this analysis, we have seen that the theorem can also
accomodate deformed Bose statistics.
One may therefore think of it as a pretty stable structure of the QFT context.
\medskip

Our identification of the
hamiltonian with the energy operator is worth emphasizing. This choice has
been motivated by the merging constraints of the time evolution of operators
in the Heisenberg picture and the cyclicity of the trace prescription. It is
in contra-distinction with what is usually defined as the hamiltonian
(eq.(2.14)), and which, we think, would lead to serious difficulties in
defining
any reliable thermal average. Exactly for the same reason, but at $T=0$, it is
not entirely obvious to us that this latter hamiltonian should be used in
order to find some small departures from a pure bosonic behavior for atoms (or
small violations of the Pauli exclusion principle, as the same strategy is
applied in that case too) [23,24]. On the other hand, if our choice is
legitimate, we have seen that the effects of the deformation (the
$\theta$-terms) comes in the thermal part of the Green's functions only, and
could therefore be difficult to observe. Nevertheless, they would come out as
pure collective effects, quite in line with some of the ideas recently
developed in [31]. \bigskip\bigskip

{\centerline{\bf Acknowledgements}} \bigskip
One of us (T.~G.), is indebted to J.~Rubin for many interesting discussions
on the mathematical aspects of the deformed algebras, and to M.~Le~Bellac
for a careful reading of the manuscript. A most motivating discussion
with H.~M.~Fried is also acknowledged. Special thanks are due to
R.~D.~Pisarski who provided the impulse for this analysis, and many fuitful
comments during its completion. T.~A. would like to thank R.~Stora for some
pertinent questions and remarks.

\bigskip\bigskip
\centerline {\bf REFERENCES}
\vskip1.5truecm

{\bf 1-} N.~P.~Landsman, {\it Nucl. Phys. A525 (1991) 397},
eds. J.P.Blaizot et. al.
\smallskip
{\bf 2-} N.~P.~Landsman and C.~G.~van~Weert, {\it Phys. Rep. 145 (1987) 141}
\smallskip
{\bf 3-} E.~Braaten and R.~Pisarski, {\it Nucl. Phys. B339 (1990) 310}
\smallskip
{\bf 4-} T.~Altherr, T.~Grandou and R.~Pisarski, {\it Phys. Lett. B271
(1991) 183}
\smallskip
{\bf 5-} T.~Muta, {\it Foundations of Quantum Chromodynamics,
(World Scientific, 1987)}
\smallskip
{\bf 6-} H.~A.~Weldon, {\it Phys. Rev. D44 (1991) 3955}
\smallskip
{\bf 7-}
R.~Baier, B.~Pire and D.~Schiff, {\it Phys. Rev. D38 (1988) 2814};
T.~Altherr, P.~Aurenche and T.~Becherrawy, {\it Nucl. Phys. B315 (1989) 436};
J.~Cleymans and D.~Dadic, {\it Z. Phys. C42 (1989) 133};
T.~Grandou, M.~Le~Bellac and J.~L~Meunier, {\it Z. Phys. C43 (1989) 575};
T.~Altherr and T.~Becherrawy, {\it Nucl. Phys. B330 (1990) 174};
T.~Altherr and P.~Aurenche, {\it Z. Phys. C45 (1990) 99};
Y.~Gabellini, T.~Grandou and D.~Poizat, {\it Ann. Phys. (N.Y) 202 (1990) 436}
\smallskip
{\bf 8-} T.~Grandou, M.~Le~Bellac and D.~Poizat, {\it Phys. Lett. B249
(1990) 478};
T.~Grandou, M.~Le~Bellac and D.~Poizat, {\it Nucl. Phys. B358 (1991) 408};
M.~Le~Bellac and P.~Reynaud, {\it Nucl. Phys. B380 (1992) 423}
\smallskip
{\bf 9-} T.~Altherr, {\it Phys. Lett. B262 (1991) 314}
\smallskip
{\bf 10-} T.~Altherr, {\it These pr\'esent\'ee \`a l'Universit\'e de Savoie,
1989}
\smallskip
{\bf 11-} J.~L.~Meunier, {\it private communications}
\smallskip
{\bf 12-} T.~Grandou, {\it Seminar given at Brookhaven, 10/12/90;
private communications}, and T.~Altherr in ref.{\bf 9}
\smallskip
{\bf 13-} G.'t~Hooft and M.~Veltman, {\it Diagrammar, CERN yellow
report $n^0$ 73-9}
\smallskip
{\bf 14-} R.~Kobes and G.~Semenoff, {\it Nucl. Phys. B260 (1985) 714;
B272 (1986) 329}
\smallskip
{\bf 15-} N.~P.~Landsman, {\it Ann. Phys. (NY) 186 (1988) 141}
\smallskip
{\bf 16-} H.~S.~Green, {\it Phys. Rev. 90 (1953) 270}
\smallskip
{\bf 17-} Y.~Ohnuki and S.~Kamefuchi, {\it Quantum Field Theory and
Parastatistics, (Springer Verlag, Berlin, 1982)}
\smallskip
{\bf 18-} A.~J~Macfarlane, {\it J. Phys. A: Math. Gen. 22 (1989) 4581}
\smallskip
{\bf 19-} R.~Mohapatra, {\it Phys. Lett. B242 (1990) 407}
\smallskip
{\bf 20-} However, one can also consider deformations about the
fermionic statistic by considering the algebra $aa^+ + qa^+a = q^{-N}$.
See, $e.g.$ J.~Beckers and N.~Debergh, {\it J. Phys. A: Math. Gen. 24
(1991) L1277}
\smallskip
{\bf 21-} E.~Floratos in {\it Quantum Groups, Proc. Argonne
Workshop, Curtright, Fairlie and Zachos (Eds), (World Scientific, 1990)}
\smallskip
{\bf 22-} O.~W.~Greenberg, {\it Phys. Rev. D43 (1991) 4111}
\smallskip
{\bf 23-} M.~L.~Ge and G.~Su, {\it J. Phys. A: Math. Gen. 24 (1991) L721}
\smallskip
{\bf 24-} M.~Kibbler and T.~Negadi, {\it Preprint LYCEN 9121}
\smallskip
{\bf 25-} S.~N.~Biswas and A.~Das, {\it Mod. Phys. Lett. 3A (1988) 549}
\smallskip
{\bf 26-} see, $e.g.$, R.~Mills, {\it Propagators for Many-Particle
Systems, (Gordon and Breach, 1969)}
\smallskip
{\bf 27-} K.~Fredenhagen, {\it Commun. Math. Phys. 79 (1981) 141}
\smallskip
{\bf 28-} L.~Baulieu and E.~Floratos, {\it Phys. Lett. B258 (1991) 171}
\smallskip
{\bf 29} J.~L.~Birman, {\it Phys. Lett. A167 (1992) 363}
\smallskip
{\bf 30-} T.~Altherr and T.~Grandou, work in progress
\smallskip
{\bf 31-} F.~Wilczek, {\it Fractional Statistics and Anyon
Superconductivity (World Scientific, 1990)}
\smallskip

\vfill\eject
{\centerline{\bf FIGURE CAPTIONS}}

\bigskip\noindent
{\bf Figure 1 :} Contour in the time complex-plane for the real
time formalism.

\smallskip\noindent
{\bf Figure 2 :} Micro-reversibility (trilinear couplings). A symbolic
representation of relations (4.5) (the diagrams stand for processes, not for
Feynman amplitudes).\smallskip\noindent {\bf Figure 3 :} The damping rate
$\Gamma_\theta(q)$ of a Higgs particle (dashed line) as given by the
imaginary part of its one-loop (dark blob) corrected two-point function.

\end